\documentclass[aps,preprint]{revtex4}%
\usepackage{amsfonts}
\usepackage{amsmath}
\usepackage{amssymb}
\usepackage{graphicx}%
\begin{document}
\title{{\LARGE ARE \ MASS AND LENGTH QUANTIZED?}}
\author{Marcelo Samuel Berman$^{1}$}
\affiliation{$^{1}$Instituto Albert Einstein\ - Av. Candido Hartmann, 575 - \ \# 17}
\affiliation{80730-440 - Curitiba - PR - Brazil}
\date{29 August, 2007.}

\begin{abstract}
We suggest that there are time-varying quanta of mass (\textit{gomidia}) and
of length (\textit{somia}), thus pointing to a quantization of geometry and
gravitation. The present numerical value of the \ \textit{gomidium}\ \ and
\ \textit{somium}\ , are, $10^{-65}$\ grams, and $10^{-91}$ centimeters.
\textit{Gomidia }may be responsible for dark matter in the Universe;
Heisenberg's principle, confirms the numerical estimates for
\ g\textit{omidia}\ \ and \ \ \ \textit{somia}\ , either for the present
Universe, or for Planck's time.

\end{abstract}
\maketitle

{\LARGE ARE \ MASS AND LENGTH QUANTIZED?}

\ \ {\large \ \ MARCELO SAMUEL BERMAN }

\bigskip

\bigskip{\large I. Introduction}

{\large \bigskip}\bigskip We introduce the definitions of micromass and
macromass, as well as those of microlength and macrolength, in the spirit of
Wesson's suggestions (Wesson, 2006). We show that by obtaining such quantities
for Planck's time, and the present Universe, \ both "micros" coincide with
Planck's mass and length, while for the present Universe, macrolength stands
as the radius of the causal Universe, while macromass represents the mass of
the Universe. We find a quantum of mass ("\textit{gomidium}") (Berman, 2007;
2007a), and a quantum of length ("\textit{somium}"), to which we suggest
interpretations. In the end of this paper, we discuss the novelties which
appear here, in comparison with what has been already published (for instance,
by Wesson, 2006). The definitions of macromass, micromass, macrolength, and
microlength, given in this paper, are related with gauge parametrizations in
penta-dimensional physics (Wesson, 2006).

\bigskip

Dark matter in the Universe, responds for 27\% of the total energy density,
which is to be represented by the critical one, as far as we accept
inflationary scenario (G\"{u}th, 1981). So we call the dark matter energy
density, \ $\rho_{\nu}$\ \ , and we write,\ 

\bigskip

$\rho_{\nu}=0.27\rho_{crit}$ \ \ \ \ \ . \ \ \ \ \ \ \ \ \ \ \ \ \ \ \ \ \ \ \ \ \ \ \ \ \ \ \ \ \ \ \ \ \ \ \ \ \ \ \ \ \ \ \ \ \ \ \ \ \ \ \ \ \ \ \ \ \ \ \ \ \ \ \ \ \ \ \ \ (1)

\bigskip

Berman (2006 b) along with others (see Sabbata and Sivaram, 1994) have
estimated that the Universe possess a magnetic field which, for Planck's
Universe, was as huge as \ $10^{55}$\ Gauss. The relic magnetic field of the
present Universe is estimated in \ $10^{-6}$\ Gauss. We can then, suppose that
some hypothetical \ particles with elementary spin, have been aligned\ \ with
the magnetic field. On the other hand, the spin of the Universe is believed to
have increased in accordance with a Machian relation. If we call
\ \ $n$\ \ \ the number of \ \textit{gomidia } in the present Universe, and
\ $n_{Pl}$\ \ its value for Planck's Universe, we may write, along with Berman
(2007; 2007a),

\bigskip

$\frac{n}{n_{Pl}}=\frac{L}{L_{Pl}}=10^{122}$\ \ \ \ \ \ \ \ . \ \ \ \ \ \ \ \ \ \ \ \ \ \ \ \ \ \ \ \ \ \ \ \ \ \ \ \ \ \ \ \ \ \ \ \ \ \ \ \ \ \ \ \ \ \ \ \ \ \ \ \ \ \ \ \ \ \ \ \ \ (2)

\bigskip

Then, \ \ \ 

\bigskip

$n=n_{Pl}\left[  \frac{R}{R_{Pl}}\right]  ^{2}$ \ \ \ \ \ \ \ \ \ \ \ \ \ . \ \ \ \ \ \ \ \ \ \ \ \ \ \ \ \ \ \ \ \ \ \ \ \ \ \ \ \ \ \ \ \ \ \ \ \ \ \ \ \ \ \ \ \ \ \ \ \ \ \ \ \ \ \ \ \ \ \ \ \ \ \ (3)

\bigskip

Thus, \ \ $n$\ \ \ grows with \ \ $R^{2}$\ .\ \ 

\bigskip

Now, we write the energy density of \ \textit{gomidia},

\bigskip

$\rho_{\nu}\cong\frac{nm_{\nu}}{\frac{4}{3}\pi R^{3}}$\ \ \ \ \ \ \ \ \ \ \ \ \ \ \ ,\ \ \ \ \ \ \ \ \ \ \ \ \ \ \ \ \ \ \ \ \ \ \ \ \ \ \ \ \ \ \ \ \ \ \ \ \ \ \ \ \ \ \ \ \ \ \ \ \ \ \ \ \ \ \ \ \ \ \ \ \ \ \ \ \ \ \ \ (4)

\bigskip

where \ \ \ $m_{\nu}$\ \ is the rest mass of the individual
\ \textit{gomidium}.

\bigskip

Berman's suggestion, in the above citations, imply that all energy densities
in the Universe decrease with \ $R^{-2}$\ . Consider for instance, the
inertial mass content. Its energy density is given by,

\bigskip

$\rho_{i}=\frac{M}{\frac{4}{3}\pi R^{3}}$%
\ \ \ \ \ \ \ \ \ \ \ \ \ \ \ \ \ \ \ \ . \ \ \ \ \ \ \ \ \ \ \ \ \ \ \ \ \ \ \ \ \ \ \ \ \ \ \ \ \ \ \ \ \ \ \ \ \ \ \ \ \ \ \ \ \ \ \ \ \ \ \ \ \ \ \ \ \ \ \ \ (5)

\bigskip

\bigskip For a Machian Universe, then, \ $M\propto R$\ \ , from where the
\ $R^{-2}$\ \ inertial energy density appears in (5).\ \ 

\bigskip

\bigskip If, then, \ \ $\rho_{\nu}\propto R^{-2}$\ \ , we find:\ \ 

1$^{st}.$) \ \ $\rho_{\nu}=0.27\rho_{Pl}\left[  \frac{R}{R_{Pl}}\right]
^{-2}$\ \ \ \ \ \ \ . \ \ \ \ \ \ \ \ \ \ \ \ \ \ \ \ \ \ \ \ \ \ \ \ \ \ \ \ \ \ \ \ \ \ \ \ \ \ \ \ \ \ \ \ \ \ \ \ \ (6)

\bigskip

2$^{nd}.$) \ \ $m_{\nu}=\frac{\rho_{Pl}R_{Pl}^{4}}{R}$\ \ \ \ \ \ \ \ . \ \ \ \ \ \ \ \ \ \ \ \ \ \ \ \ \ \ \ \ \ \ \ \ \ \ \ \ \ \ \ \ \ \ \ \ \ \ \ \ \ \ \ \ \ \ \ \ \ \ \ \ \ \ \ \ \ \ \ \ \ \ \ (7)

\bigskip

We see now that while the number of \ \ \textit{gomidia \ }in the Universe
increases with \ $R^{2}$\ , the rest mass\ decreases with \ \ $R^{-1}$\ \ ; we
may obtain, with \ \ $R\cong10^{28}$\ cm, that the rest mass of \textit{
\ gomidia } should be, in the present Universe:

\bigskip

$m_{\nu}\cong10^{-65}$\ g\ \ \ \ \ \ \ \ \ \ .\ \ \ \ \ \ \ \ \ \ \ \ \ \ \ \ \ \ \ \ \ \ \ \ \ \ \ \ \ \ \ \ \ \ \ \ \ \ \ \ \ \ \ \ \ \ \ \ \ \ \ \ \ \ \ \ \ \ \ \ \ \ \ \ \ \ \ \ \ \ \ (8)

\bigskip

\textbf{\bigskip}A law of variation for the number of  \textit{gomidia }\ in
the Universe has been found. A law of variation for the rest mass of
\textit{gomidia}\ was also found.

\bigskip

We remind the reader that Kaluza-Klein's cosmology (Wesson, 1999; 2006; Berman
and Som, 1993), considers time varying rest masses, in a penta-dimensional
("induced mass") space-time-matter, of which the fifth coordinate is rest
mass. The above results can not be rejected, for the time being, by any known
data. \ \ \ We point out, that some of the features of the present
\ calculation, resemble some points in a paper by Sabbata and Gasperini (1979).

\bigskip

{\large II. Quantization of inertia and geometry}

\bigskip

\ \ \ Wesson (2006), by citing Desloge(1984), comments that by means of the
four fundamental "constants", Planck's ($h$), Newton's ($G$), speed of light
($c$), and cosmological ($\Lambda$), one can obtain two different kind of
mass, the micromass ($m$), and the macromass ($M$), given by:

\bigskip

$m=\left(  \frac{h}{c}\right)  \Lambda^{1/2}$
\ \ \ \ \ \ \ \ \ \ \ \ \ \ \ \ , \ \ \ \ \ \ \ \ \ \ \ \ \ \ \ \ \ \ \ \ \ \ \ \ \ \ \ \ \ \ \ \ \ \ \ \ \ \ \ \ \ \ \ \ \ \ \ \ \ \ \ \ \ (9)

\bigskip

and,

\bigskip

$M=\frac{c^{2}}{G}\Lambda^{-1/2}$ \ \ \ \ \ \ \ \ \ \ \ \ \ \ \ \ . \ \ \ \ \ \ \ \ \ \ \ \ \ \ \ \ \ \ \ \ \ \ \ \ \ \ \ \ \ \ \ \ \ \ \ \ \ \ \ \ \ \ \ \ \ \ \ \ \ \ \ \ \ \ \ (10)

\bigskip

\bigskip Micromass involves Planck's constant, hence its denomination;
macromass is defined by means of \ $G$\ \ , so its "macro" denomination.

\bigskip Notice that the above constant tetrad, is, of course, overlapping.
With the present values for the cosmological "constant", \ $\Lambda
=\Lambda_{U}\approx10^{-56}$\ cm$^{-2}$\ , it is found,

\bigskip

$m_{(U)}\approx10^{-65}$ g \ \ \ \ \ \ \ , \ \ \ \ \ \ \ \ \ \ \ \ \ \ \ \ \ \ \ \ \ \ \ \ \ \ \ \ \ \ \ \ \ \ \ \ \ \ \ \ \ \ \ \ \ \ \ \ \ \ \ \ \ \ \ \ \ \ \ \ \ (11)

\bigskip

and,

\bigskip

$M_{(U)}\approx10^{56}$ \ g \ \ \ \ \ \ \ . \ \ \ \ \ \ \ \ \ \ \ \ \ \ \ \ \ \ \ \ \ \ \ \ \ \ \ \ \ \ \ \ \ \ \ \ \ \ \ \ \ \ \ \ \ \ \ \ \ \ \ \ \ \ \ \ \ \ \ \ \ (12)

\bigskip

The present Universe's micromass ($m_{(U)}$), represents a present
mass-quantum, i.e., the minimum mass in the present Universe. On the other
hand, the present Universe's macromass ($M_{(U)}$), is approximately the mass
of the present Universe ($M_{U}$) .

\bigskip

What Wesson overlooked, is that, when we apply the definitions (9) and (10),
\ by plugging, \ $\Lambda=\Lambda_{PL}\approx L_{PL}^{-2}\approx10^{-66}%
$\ cm$^{-2}$\ , which stand for Planck's time values, we find that micromass
and macromass coincide approximately with Planck's mass, $M_{PL}$\ , \ \ 

\bigskip

$M_{(PL)}=m_{(PL)}=M_{PL}\approx10^{-5}$ g \ \ . \ \ \ \ \ \ \ \ \ \ \ \ \ \ \ \ \ \ \ \ \ \ \ \ \ \ \ \ \ \ \ \ \ \ \ \ \ \ \ \ \ \ \ \ (13)

\bigskip

We are led to consider that, macromass, is always associated to the mass of
the Universe ($M_{U}$), either in the very early Universe or in the present one.

\bigskip As to the micromass, we baptize this mass as the quantum mass value
(\textit{gomidium, }after F.M.Gomide): it is a time-varying mass, because
\ \ $\Lambda$\ \ is also so, because we expect its energy density
\ \ $\frac{\Lambda}{\kappa}$\ \ depend on \ \ $R^{-2}$\ \ altogether.\ \ \ 

\bigskip

We now show, that associated with micromass and macromass, we have two
distinct length values, which come associated to the present and Planck's Universe.

\bigskip

For each mass, we associate two kinds of lengths, namely, the macrolength,
($\lambda_{\nu}$), and the microlength ($l_{\nu}$); the first one, is
Compton's wavelength, \ given by,

\bigskip

$\lambda_{\nu}=\frac{\bar{h}}{mc}$ \ \ \ \ \ \ \ \ \ \ \ \ \ \ \ . \ \ \ \ \ \ \ \ \ \ \ \ \ \ \ \ \ \ \ \ \ \ \ \ \ \ \ \ \ \ \ \ \ \ \ \ \ \ \ \ \ \ \ \ \ \ \ \ \ \ \ \ \ \ \ (14)

\bigskip

\bigskip It is a "macro", because it is inversely proportional to micromass.

The second length, that we call "microlength", is a gravitationally associated
length with mass, which we term the quantum of length, or \textit{somium} ,

\bigskip

$l_{\nu}=\frac{Gm}{c^{2}}$ \ \ \ \ \ \ \ \ \ \ \ \ \ . \ \ \ \ \ \ \ \ \ \ \ \ \ \ \ \ \ \ \ \ \ \ \ \ \ \ \ \ \ \ \ \ \ \ \ \ \ \ \ \ \ \ \ \ \ \ \ \ \ \ \ \ \ \ \ \ \ \ (15)

\bigskip

\bigskip It is a "micro", because it is proportional to micromass.

We find, a microlength, $l_{\nu}$ ,for the present Universe, with
\ \ \ $m=m_{\nu}=m_{(U)}$ \ \ ,

\bigskip

$l_{\nu(U)}\approx10^{-91}$ \ \ cm, \ \ \ \ \ \ \ \ \ \ \ \ \ \ \ \ \ \ \ \ \ \ \ \ \ \ \ \ \ \ \ \ \ \ \ \ \ \ \ \ \ \ \ \ \ \ \ \ \ \ \ \ \ \ \ \ \ \ (16)

\bigskip

while, \ the macrolength is then found,

\bigskip

$\lambda_{\nu(U)}\approx10^{28}$ cm \ \ \ \ . \ \ \ \ \ \ \ \ \ \ \ \ \ \ \ \ \ \ \ \ \ \ \ \ \ \ \ \ \ \ \ \ \ \ \ \ \ \ \ \ \ \ \ \ \ \ \ \ \ \ \ \ \ \ \ \ \ (17)

\bigskip

On the other hand, for Planck's Universe, the macrolength is found to be,

\bigskip

$\lambda_{\nu(PL)}\approx10^{-33}$ cm \ \ \ \ ,\ \ \ \ \ \ \ \ \ \ \ \ \ \ \ \ \ \ \ \ \ \ \ \ \ \ \ \ \ \ \ \ \ \ \ \ \ \ \ \ \ \ \ \ \ \ \ \ \ \ \ \ \ \ \ (18)

\bigskip

and, the microlength has the same numerical value,

\bigskip

$l_{\nu(PL)}\approx10^{-33}$ cm \ \ . \ \ \ \ \ \ \ \ \ \ \ \ \ \ \ \ \ \ \ \ \ \ \ \ \ \ \ \ \ \ \ \ \ \ \ \ \ \ \ \ \ \ \ \ \ \ \ \ \ \ \ \ \ \ \ \ \ \ (19)

\bigskip

One can check, that the microlength represents a quantum, the \textit{somium.}
Macrolength is represented by the radius of the Universe.

\bigskip

For Planck's Universe, the \textit{somium,} the macrolength, and the
microlength then coincide with the Planck's radius, $L_{PL}$\ .\ \ 

\bigskip

\bigskip{\large III. Heisenberg uncertainty and minimum mass and length}

\bigskip

\bigskip According to Heisenberg's uncertainty principle, any two conjugate
quantities, in the sense of Hamilton's canonical ones, carry uncertainties,
\ $\Delta Q$\ \ and \ \ $\Delta P$\ \ , which obey the condition (Leighton,
1959), 

\bigskip

$\Delta Q\Delta P\approx h$\ \ \ \ \ \ \ \ \ \ \ \ \ \ \ \ \ .\ \ \ \ \ \ \ \ \ \ \ \ \ \ \ \ \ \ \ \ \ \ \ \ \ \ \ \ \ \ \ \ \ \ \ \ \ \ \ \ \ \ \ \ \ \ \ \ \ \ \ \ (20)

\bigskip

If we consider maxima \ $\Delta P$\ , we obtain minima \ $\Delta Q$\ \ .

\bigskip

If \ $\Delta P$\ \ stands for the uncertainty in linear momentum, given, say,
by the product of mass and speed, then, its maximum value must be the product
of the largest mass in the Universe by the largest speed, 

\bigskip

$\Delta P=M_{U}$ $c$ \ \ \ \ \ \ \ \ \ \ \ \ \ \ . \ \ \ \ \ \ \ \ \ \ \ \ \ \ \ \ \ \ \ \ \ \ \ \ \ \ \ \ \ \ \ \ \ \ \ \ \ \ \ \ \ \ \ \ \ \ \ \ \ \ \ \ \ \ \ \ (21)

\bigskip

We thus, obtain a minimum length value, 

\bigskip

$\Delta Q\approx\frac{h}{c\text{ }M_{U}\text{ }}$
\ \ \ \ \ \ \ \ \ \ \ \ \ \ \ \ . \ \ \ \ \ \ \ \ \ \ \ \ \ \ \ \ \ \ \ \ \ \ \ \ \ \ \ \ \ \ \ \ \ \ \ \ \ \ \ \ \ \ \ \ \ \ \ \ \ \ \ \ \ (22)

\bigskip

Now, let us think of the largest time numerical value in the Universe, 

\bigskip

$t_{U}\approx10^{10}$ \ \ years. \ \ \ \ \ \ \ \ \ \ \ \ \ \ \ \ \ \ \ \ \ \ \ \ \ \ \ \ \ \ \ \ \ \ \ \ \ \ \ \ \ \ \ \ \ \ \ \ \ \ \ \ \ \ \ \ \ \ \ \ \ \ \ \ (23)

\bigskip

Its conjugate variable, will point out to a minimum inertial energy and a
minimum inertial mass ( $\Delta m$\ )\ \ \ , \ \ 

$\bigskip$

$\Delta E=c^{2}\Delta m\approx\frac{h}{t_{U}}$\ \ \ \ \ \ \ \ \ \ \ \ \ \ \ . \ \ \ \ \ \ \ \ \ \ \ \ \ \ \ \ \ \ \ \ \ \ \ \ \ \ \ \ \ \ \ \ \ \ \ \ \ \ \ \ \ \ \ \ \ \ \ \ (24)

\bigskip

It turns out, that we have retrieved, from \ $\Delta m$\ \ , and \ \ $\Delta
Q$\ \ , the minimum mass and length for the present Universe, with the same
approximate values attached to \ \textit{gomidia}\ \ and \ \ \textit{somia}
\ , in last Section,\ also for the present Universe.

\bigskip

Analogously, we could repeat the calculation for Planck's time and Planck's
mass, and we then would obtain numerically the same values attained by
\ \ \textit{gomidia}\ \ and \ \ \textit{somia} \ in Planck's Universe.\ \ \ 

\bigskip

We have, then, support for the quantization of mass and length, in a
time-varying fashion, coinciding with the calculation in the last section. \ \ \ \ 

\bigskip

\bigskip{\large IV. Conclusions}

\bigskip

\bigskip Wesson (2006), dealt classically with \ $D=5$\ \ WEP (weak
equivalence principle), as a symmetry in the "induced
mass"\ \ pentadimensional Kaluza-Klein theory, or even the brane one. The new
fifth forces and coordinate are then present. The geodesic equation adds and extra-acceleration.

\bigskip

Microlength and macrolength were found by Wesson, to be good gauge
parametrizations for mass, which allow a mass geometry consistent with the
rest of Physics. The known laws of Mechanics and conservation of linear
momentum, in limiting cases, were also preserved when \ $m$\ \ is a
representation of rest-mass. Because of the standard structure of
\ \ $D=5$\ \ Physics as an extension of the \ $D=4$\ \ chapter, while
introducing an extra coordinate, opens the possibility of quantization in the
lower dimension Physics. Wesson even advanced that the Quantum domain would
extend to the Cosmos, in the form of a broken symmetry for the angular momenta
tied to the gravitational field.

We have therefore, found in our present paper, that the micromass and
microlength represent quanta of mass and length. We call them, respectively,
\textit{gomidium} and \textit{somium}, but their numerical values are
time-varying: present day's \ \textit{gomidium}\ is 10$^{-65}$\ g, while
\ \textit{somium}\ is about 10$^{-91}$\ cm. Planck's values for
\textit{gomidium} and \textit{somium}, coincide respectively with Planck's
mass and Planck's length.\ \ We have thus hinted that mass is quantized, but
geometry is altogether. As gravitation is associated with geometry,
quantization of the latter, implies on the former: it seems that quantum
gravity has been found. Much of what we have calculated here, like the Machian
derivation, which led to time-varying quanta of mass and length, and also the
interpretation, under which macromass and macrolength describe the Universe's
mass and radius, throughout its lifespan, (in particular, Planck's and present
times) are novelties in the literature. Though some of the topics dealt in our
paper, were sparsely dealt in Wesson's books (Wesson, 1999; 2006), and by
other authors, we have here given a rational interpretation of otherwise
disconnected elements. The numerical value for the present Universe's
microlength (\textit{somium}), seems to have never appeared in the literature,
in any context; \ our quantization ideas, as far as I know, are also novel;
dark matter has been associated with \ \textit{gomidia}. All the above,
supported by Heisenberg's uncertainty principle.\ 

\bigskip

\bigskip{\Large Acknowledgments}

\bigskip

The author expresses his recognition to his intellectual mentors, now friends
and colleagues, M.M. Som and F.M. Gomide. He thanks the many other colleagues
that collaborate with him, and the Editor of this Journal, for providing an
important objection towards an earlier manuscript. The typing was made by
Marcelo F. Guimar\~{a}es, who I consider a friend and to whom my thanks are
due for this and many other collaborations.

\bigskip\bigskip

{\Large References}

\bigskip

\bigskip Berman,M.S. (2006) - \textit{Energy of Black-Holes and Hawking's
Universe \ }in \textit{Trends in Black-Hole Research, }Chapter 5\textit{.}
Edited by Paul Kreitler, Nova Science, New York.

Berman,M.S. (2006 a) - \textit{Energy, Brief History of Black-Holes, and
Hawking's Universe }in \textit{New Developments in Black-Hole Research},
Chapter 5\textit{.} Edited by Paul Kreitler, Nova Science, New York.

Berman,M.S. (2006 b) - Los Alamos Archives, http://arxiv.org/abs/physics/0606208.

Berman,M.S. (2007) - \textit{Introduction to General Relativity, and the
Cosmological Constant Problem}, Nova Science, New York..

Berman,M.S. (2007a) - \textit{Introduction to General Relativistic and
Scalar-Tensor Cosmologies}, Nova Science, New York..

Berman,M.S.; Som, M.M. (1993) - Astrophysics and Space Science, \textbf{207}, 105.

Brans, C.; Dicke, R.H. (1961) - Physical Review, \textbf{124}, 925.

Chen,W.;Wu,Y.- S. (1990) - Phys. Review D \textbf{41},695.

Desloge, E.A. (1984) - Am. J. Phys. \textbf{52}, 312.

Gomide, F.M.(1963) - Nuovo Cimento, \textbf{30}, 672.

Guth, A. (1981) - \textit{Physical Review, }\textbf{D23,} 347.

Leighton, R.B. (1959) - \textit{Principles of Modern Physics,} McGraw-Hill, N.Y.

Sabbata, V. de; Gasperini, M. (1979) - Lettere al Nuovo Cimento. \textbf{25}, 489.

Sabbata, V. de; Sivaram, C. (1994) - \textit{Spin and Torsion in Gravitation,}
World Scientific, Singapore.

Wesson, P.S. (1999) - \textit{Space-Time-Matter}, World Scientific, Singapore.

Wesson, P.S. (2006) - \textit{Five dimensional Physics}, World Scientific, Singapore.

\end{document}